\documentclass[aps,prl,floatfix,twocolumn,footinbib,amsmath,amssymb]{revtex4-1}
\usepackage                             {epsfig}
\usepackage                             {natbib}
\usepackage                             {xspace}
\usepackage                             {amsmath}
\usepackage                             {graphicx}
\usepackage     [english]               {babel}
\usepackage                             {natbib}
\usepackage                             {hyperref}
\usepackage                             {amsfonts}
\usepackage                             {amssymb}
\usepackage                             {latexsym}
\newcommand{\figref}[1]{Fig.~\ref{#1}}
\newcommand{\figureref}[1]{Figure~\ref{#1}}

\newcommand{\eref}[1]{(\ref{#1})}
\renewcommand{\eqref}[1]{Eq.~(\ref{#1})}
\newcommand{\ket}[1]{| #1 \rangle}

\newcommand{\braket}[1]{\langle #1 \rangle}
\begin{document}
\title{Inelastic Confinement-Induced Resonances in Low-Dimensional Quantum Systems}
\author{Simon Sala, Philipp-Immanuel Schneider, and Alejandro Saenz}

       \affiliation{AG Moderne Optik, Institut f\"ur Physik,
         Humboldt-Universit\"at zu Berlin, Newtonstrasse 15,
         12489 Berlin, Germany}
       \date{\today}
\begin{abstract}
  A theoretical model is presented describing the confinement-induced
  resonances observed in the recent loss experiment of Haller et
  al. [Phys. Rev. Lett. 104, 153203 (2010)].
  These resonances originate from possible molecule formation  due to the
  coupling of center-of-mass and relative motion.  
  A corresponding model is verified by ab initio calculations and
  predicts the resonance positions in 1D as well as in 2D confinement
  in agreement with the experiment. This resolves the contradiction of
  the experimental observations to previous theoretical predictions.
\end{abstract}
\maketitle
%
%
%
Low-dimensional quantum systems show intriguing phenomena.
For example, a 2D confinement allows for the existence of
particles with fractional statistics known as anyons \cite{cold:naya08}. In 1D
a gas of impenetrable Bosons, the 
Tonks-Girardeau gas%
, acquires Fermionic properties \cite{cold:hall09,cold:pare04,*cold:kino04}.
Nowadays, one and two-dimensional systems can be experimentally
realized in trapped ultracold gases offering a high degree of control
\cite{cold:guen05,cold:hall09,cold:hall10b}.
In low-dimensional systems confinement-induced resonances (CIRs) 
have attracted great interest.
They are
universal, since they depend solely on the geometry of the trap. In
1998, Olshanii developed a mapping of the relative-motion Hamiltonian
of a system of two atoms confined in a harmonic trap at large
anisotropies onto the corresponding purely one-dimensional one
\cite{cold:olsh98}. The resulting 1D effective interaction strength
$g_{1D}$ shows a divergent behavior at a specific scattering length
which leads to the formation of a Tonks-Girardeau gas
\cite{cold:gira60,cold:lieb63}. An analogous derivation of the
effective 2D interaction strength $g_{2D}$ reveals a similar divergent
behavior \cite{cold:petr00}.
The divergences in these models, which are based on the relative-motion
Hamiltonian in harmonic approximation (RMH models), were confirmed in 1D by ab
initio calculations of $g_{1D}$ \cite{cold:berg03} and very recently in 2D
adopting radio-frequency spectroscopy to measure $g_{2D}$ \cite{cold:froh11}.
An experimental search for the predicted CIRs in terms of particle losses and
heating was performed by Haller \emph{et al.}  \cite{cold:hall10b}. 
In 1D, a resonance close to the position predicted by the RMH model was found
for isotropic transversal confinement. However, the experiment considered also
an anisotropic transversal confinement and found an unexpected splitting of
the resonance. Furthermore, in 2D confinement a resonance was observed at
positive values of the scattering length. A detailed analysis
\cite{cold:peng10,*cold:zhan11}
proved formally that these two observations contradict the RMH
models that predict 
a \emph{single} resonance in 1D and a resonance at
\emph{negative} scattering lengths in 2D confinement.
The immediate question arises what kind of resonances were observed in
\cite{cold:hall10b}.  Are the RMH resonances modified due to the
experimental setup, or are the losses in \cite{cold:hall10b} of
completely different origin? If so, why are RMH resonances not seen in
terms of losses?


In this Letter it will be demonstrated that the resonances observed in
\cite{cold:hall10b} are caused by a coupling of the center-of-mass (COM) and
relative (REL) motion which originates from the anharmonic terms of the
trapping optical lattice. The COM-REL coupling (CRC) and strong 1D or 2D
confinement leads to Feshbach resonances induced by (avoided) crossings of
bound states with COM excitation and a state of an unbound atom pair
without COM excitation.
Hence, at the avoided crossing a molecular bound state can be occupied, since
the excess binding energy can be transferred to COM excitation energy.  In
view of this energy transfer we refer to the COM-REL resonance as an \emph{inelastic}
CIR. The formation of molecules is a basic loss mechanism and should thus be
observable in loss experiments like \cite{cold:hall10b}.
On the contrary, in the spectra of the RMH Hamiltonians
\cite{cold:busc98,*cold:idzi06} no curve crossings appear that could lead to a
significant occupation of an excited bound state and corresponding molecule
formation \footnote{The shifted bound state shown in Fig.1 in
  \cite{cold:hall10b} and Fig.2 in \cite{cold:berg03} does, in fact, not exist
  in the energy spectrum of the \emph{full} relative motion Hamiltonian, see
  corresponding figures in \cite{cold:busc98,*cold:idzi06}, but of a projected
  Hamiltonian.}. Because of the absence of an energy transfer the CIRs of the
RMH models are denoted as \emph{elastic} CIRs in the following.

The CRC model is introduced, first for 1D and later for 2D
confinement, for two identical particles in agreement with the
experiment.
The external potential is assumed to be expandable in a power series
around the origin and separable in the three spatial directions. In
COM and REL coordinates, $\mathbf{r}= \mathbf{r_1}-\mathbf{r_2}$ and
$\mathbf{R}=\frac{1}{2}(\mathbf{r_1}+\mathbf{r_2})$, the Hamiltonian
is given as
\begin{align}%
 H(\mathbf{r},\mathbf{R}) =\ & T_{\rm REL}(\mathbf{r}) + T_{\rm COM}(\mathbf{R})
+  V_{\rm REL}(\mathbf{r})\nonumber\\  
& + V_{\rm COM}(\mathbf{R})
+ U_{\rm int}(r)  + W(\mathbf{r}, \mathbf{R})
\label{eq:6tic_Hamil_full}  
\end{align}%
where $T_{\mathrm{REL}}$ and $T_{\mathrm{COM}}$ are 
the kinetic-energy operators of the REL and COM motion, respectively. $V_{\rm REL}$ and
$V_{\rm COM}$ are the separable parts of the  potential energy. Thus, $W$  contains only
 the non-separable terms that are of the form $r_i^n R_i^m$ with $i \in \{x,y,z\}$
and $n,m \in \mathbb{N}\backslash\{0\}$.
$U_{\rm int}(r)$ is the inter-particle interaction which in the CRC
model is described by the pseudo potential $U_{\rm int}(r)=\frac{4 \pi \hbar^2 a}{m}
\delta(\mathbf{{r}})\frac{\partial}{\partial r}r$ where $a$ is the 3D $s$-wave scattering 
length and $m$ the atom mass.
In the case of an optical lattice in three spatial
directions the external potential terms read
\begin{align}
  \label{eq:pot1}
  V_{\rm REL}(\mathbf{r}) = 2 \sum_{j=x,y,z} V_j \sin^2( \frac{1}{2} k  r_j)
\end{align}
\begin{align}
  \label{eq:pot2}
  V_{\rm COM}(\mathbf{R}) = 2 \sum_{j=x,y,z}V_j \sin^2(k R_j)
\end{align}
\begin{align}
  \label{eq:pot3}
  W(\mathbf{r},\mathbf{R}) = - 4 \sum_{j=x,y,z} V_j \sin^2( \frac{1}{2} k r_j) \sin^2( k  R_j)    
\end{align}
with $k=\frac{2\pi}{\lambda}$ and $\lambda$ the laser wavelength.
$V_j$ is the lattice depth in direction $j\in\{x,y,z\}$. In an optical lattice
1D geometry can be achieved by forming decoupled quasi 1D tubes which requires a 
sufficient lattice depth in two transversal directions, e.g., $x$ and $y$.
Consequently, it suffices to consider the physics of a single tube.  

To estimate the energy spectrum the harmonic approximation
of a single tube with trap frequencies $\omega_j$ and trap lengths 
$d_j = \sqrt{2 \hbar/(m \omega_j)}$ is considered. 
The  confinement is characterized by the anisotropies $\eta_x=\omega_x/\omega_z$ and
$\eta_y=\omega_y/\omega_z$, which should be sufficiently large.
Within the harmonic approximation 
the coupling between COM and REL motion vanishes. The wavefunctions are thus represented
by products of COM and REL eigenstates. 
The harmonic relative motion Hamiltonian possesses a single bound state
$\psi^{(b)}(\mathbf{r})$ \cite{cold:lian08}.
The influence of the weak longitudinal
confinement on the bound state energy $E_b^{\rm REL}(a)$ can be
neglected \cite{cold:idzi06}. 
In this case $E_b^{\rm REL}$ and $a$ satisfy the relation
 \cite{cold:peng10}
\begin{align}
\label{eq:bound_state_ener}
  \frac{\sqrt{\pi} d_y}{a}=-\int_{0}^{\infty }
\left(
  \frac{\sqrt {\beta}{{\rm e}^{\frac{t\epsilon}{2}}}}
       {\sqrt {t \left( 1-{{\rm e}^{-\beta\,t}} \right)  
        \left( 1-{{\rm e}^{-t}} \right) }}
  -t^{-\frac{3}{2}}
\right)dt
\end{align}
where $\epsilon=(E_b^{\rm REL}-E_0)/(\hbar \omega_y)$,
$E_0=\frac{\hbar}{2}(\omega_x + \omega_y)$, and $\beta=\omega_x/\omega_y$.  A
general expression for the eigenenergies above the REL motion threshold
$E_{\rm th}=\frac{\hbar}{2}(\omega_x + \omega_y +\omega_z)$ is not
known. 
The states above $E_{\rm th}$ are denoted as trap states in the
following.
The eigenenergy $E_{\rm 1}^{\rm REL}$ of the first trap state $\ket{\psi_1}$
lies in the interval $[E_{\rm th},E_{\rm th}+ 2 \hbar \omega_z)$ such that for
the model $E_{\rm 1}^{\rm REL}$ is approximated by $E_{\rm th}+\hbar \omega_z
$~\footnote{It turns out that this approximation is exact at the resonance position 
   in the case of transversal isotropic, harmonic confinement.}.
 In the limit as the temperature $T \to 0$, the higher
excitations are frozen out.
The eigenstates of the COM Hamiltonian factorize as
$\Phi_\mathbf{n}(\mathbf{R}) = {\phi_{n_x}({X})}\, {\phi_{n_y}({Y})}\,
{\phi_{n_z}({Z})} $ with $\mathbf{n}=(n_x,n_y,n_z)$ and eigenenergies
$E_\mathbf{n}^{\rm COM}=\sum_{j=x,y,z}\hbar \omega_j(n_j+\frac{1}{2}$).  When
combining REL and COM motion the energies of the bound states become $E_b^{\rm
  REL}(a)+E_\mathbf{n}^{\rm COM}$ while the energy of the lowest trap state is
given by $E_{\rm 1}^{\rm REL}+E_{(0,0,0)}^{\rm COM}$. For $n\ne (0,0,0)$
\emph{crossings} between the excited bound state and the lowest trap state
occur for
\begin{align}
 \label{eq:crossing}
  E_b^{\rm REL}=E_{\rm  1}^{\rm REL}-\Delta_\mathbf{n}  
\end{align}
where $\Delta_\mathbf{n}=E_\mathbf{n}^{\rm COM}-E_{(0,0,0)}^{\rm COM}$ is the COM excitation.
The corresponding scattering length at the crossing is obtained from 
\eqref{eq:bound_state_ener}.

In the following, the influence of the anharmonicity of the confining
potentials of Eqs.~\eref{eq:pot1}, \eref{eq:pot2}, and \eref{eq:pot3} is
considered.  First of all, the non-vanishing term
$W(\mathbf{r},\mathbf{R})\neq 0$ leads to the coupling between REL and
COM states of equal symmetry. 
Resonances are only observable if the coupling between the
crossing states is sufficiently strong.  
Within the CRC model, the matrix elements defining the coupling strength between a
bound state $\ket{\Phi_\mathbf{n}\, \psi^{(b)}}$ with COM excitation
$\Delta_{\mathbf{n}}$ and the lowest trap state $\ket{\Phi_{(0,0,0)}\, \psi_1}$
are $ W_\mathbf{n}= \braket{\Phi_\mathbf{n}\, \psi^{(b)} | W |
  \psi_1\, \Phi_{(0,0,0)}}$.
Since $W(\mathbf{r},\mathbf{R}) = \sum_{j=x,y,z} W_j({r_j},{R_j})$
separates in the spatial directions the matrix elements become
\begin{align}
\label{eq:matrix_ele}
& W_\mathbf{n} = \sum_{j=x,y,z} \braket{\Phi_\mathbf{n}\, \psi^{(b)}| W_j | \Phi_{(0,0,0)}\, \psi_1}.
\end{align}
The coupling to a bound state along the weakly confined $z$ direction 
can be neglected, because the bound state has 
an extension $d_b \ll d_z$.
Hence, $W(z,Z)\approx W(0,Z)$ holds within the extension of the
bound state $|z| \leq d_b$ and due to the orthogonality of the REL
eigenstates the longitudinal matrix element vanishes,
$  \braket{\Phi_\mathbf{n} \psi^{(b)}| W_z | \psi_1 \Phi_{(0,0,0)} }
 \approx  \braket{ \Phi_\mathbf{n} | W_z(z=0,Z) | \Phi_{(0,0,0)}}
\braket{ \psi^{(b)} | \psi_1 }=0$.
With this approximation the coupling matrix element \eref{eq:matrix_ele} becomes
\begin{align}
  \label{eq:final_select}
   W_\mathbf{n} \approx  \delta_{n_z,0} 
   \times \bigg[ 
   &\delta_{n_y,0} \braket{\phi_{n_x} \psi^{(b)} | W_x | \psi_1 \phi_{0} }\nonumber\\ 
  &+\delta_{n_x,0} \braket{\phi_{n_y} \psi^{(b)} | W_y | \psi_1 \phi_{0} } 
   \bigg].
\end{align}
As implied by \eqref{eq:crossing}, the excitation of the bound state
$\Delta_{\mathbf{n}}$ must be nonzero for crossings to exist.  It
follows directly from \eqref{eq:final_select} that resonances can only
occur for excitations of the bound state in a \emph{single
  transversal} direction. For symmetry reasons the
excitations must be even.  The couplings connected to the lowest
transversal excitations are dominant, because for higher excitations
the increasing oscillatory behavior of the wavefunctions reduces the
values of the integrals in \eqref{eq:final_select}. Hence, the
inelastic
CIRs for the quasi 1D confinement arise dominantly from the coupling
matrix elements with quantum numbers $n=(2,0,0)$ and $n=(0,2,0)$.

Based on these selection rules 
the positions of the inelastic CIR can be predicted.
So far, the harmonic approximation is used to
determine the energy crossings. For the excitations of the bound state higher
COM states are involved so that the influence of the anharmonicity on the COM
energy cannot be neglected. Therefore, $\Delta_\mathbf{n}$ in
\eqref{eq:crossing} is corrected within first-order perturbation theory
treating the leading anharmonic (quartic) term of the COM optical lattice, $-
\frac{1}{24} \sum_{j=x,y,z} \frac{\hbar \omega_j}{V_j} R_j^4$, as a
perturbation. The resulting energy offset $\Delta_\mathbf{n}$ becomes
\begin{align}
\label{eq:Delta}
  \Delta_{(n_x,n_y,n_z)}  =\sum_{j=x,y,z} \hbar \omega_j n_j \left(
      1 - \frac{\hbar \omega_{j}}{16 V_j} 
       \left[ n_j+1 \right] \right).
\end{align}

\begin{figure}[ht]
  \begin{centering}
  \includegraphics[width=0.42\textwidth]{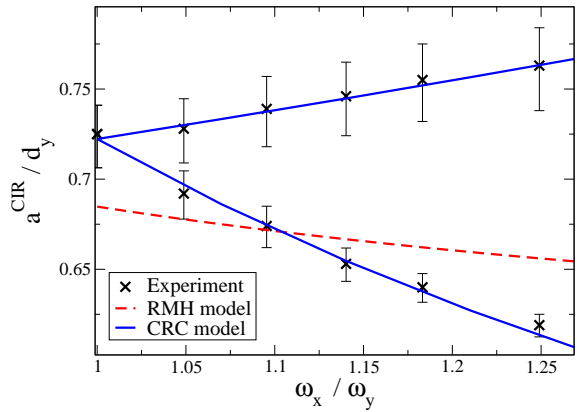}
  \caption{(color online) Positions of CIRs in terms of the scattering length for different
    values of transversal anisotropy in quasi 1D. The experimental positions
    of maximal particle loss are compared to the predicted CIR positions of
    the RMH model and the CRC model using the experimental parameters (Cs atoms confined in a trap with $\eta_y =
    825$, $V_y = 24.8 E_r$ and $\lambda=1064.49 \ {\rm nm}$ where $E_r = \hbar^2 k^2 / 2m $ is the photon recoil energy.}
  \label{fig:cir_splitting_num_formula}
  \end{centering}
\end{figure}

The CIR positions can now be easily determined using
Eqs.~\eref{eq:crossing} and \eref{eq:Delta} with $n=(2,0,0)$ and
$n=(0,2,0)$. \figureref{fig:cir_splitting_num_formula} shows that the
predicted resonance positions agree perfectly to the experimental
positions of maximal atom loss. Note, while in \cite{cold:hall10b} the
``edges'', i.e. the scattering length for which the atom loss starts
to rise significantly, were chosen as resonances positions, in the CRC
model the conventional positions of maximal particle loss are
considered \footnote{The maximal loss positions are determined by
  shifting the known ``edge'' positions by a constant offset $\Delta a
  = 89\ a_0$ that is obtained from the isotropic case.}.
The CRC model explains accurately the observed splitting of the
resonance for anisotropic transversal confinement.
%

\begin{figure}[ht]
  \begin{centering}
    \includegraphics[width=0.42\textwidth]{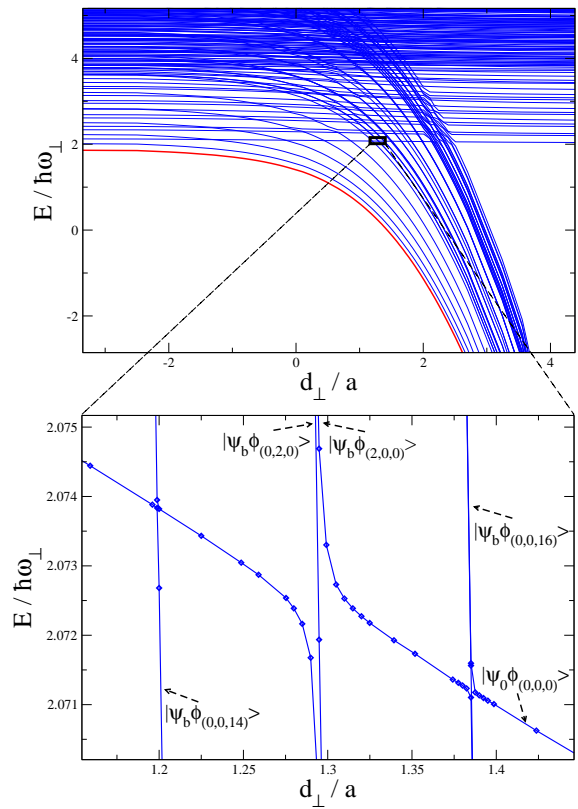}
    \caption{(color online) Spectrum of the full coupled Hamiltonian
      \eref{eq:6tic_Hamil_full} for $^7$Li atoms confined in a sextic
      trapping potential with $ V_x = V_y = 35.9 E_r $,
      $\eta_x=\eta_y=10$ and $\lambda=1000 \ {\rm nm}$. In the upper
      part all states bending down to $-\infty$ are molecular states
      originating from the REL bound state $\psi^{(b)}$ with different COM
      excitations. The one in the COM ground state is marked red. The
      magnified part shows the avoided crossing responsible for the
      inelastic CIR which arises from the crossings of the
      transversally excited bound states with the ground trap
      state. The kets indicate the dominant contribution to the
      states. Only transversally excited states couple strongly while
      for longitudinal excitation the coupling is very small resulting
      in almost non-avoided crossings.  For isotropic transversal
      confinement only one CIR occurs due to the degeneracy of the
      transversal excitation.}
    \label{fig:spectrumplot_ci_sextic_ix5000_iy5000_iz50_zoom2}
  \end{centering}
\end{figure}

\begin{figure}[ht]
  \begin{centering}
    \includegraphics[width=0.42\textwidth]{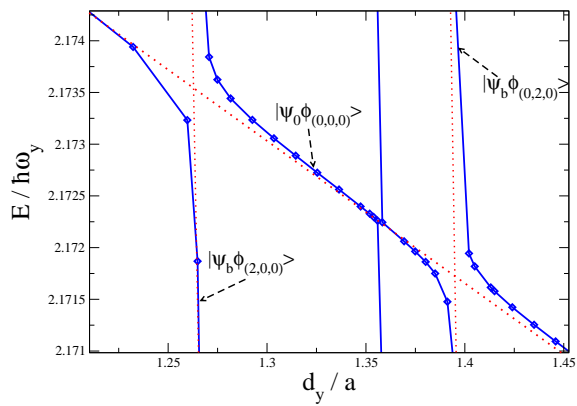}
    \caption{(color online) Spectrum of the full coupled Hamiltonian
      \eref{eq:6tic_Hamil_full} for $^7$Li atoms confined in a sextic
      trapping potential with $V_y = 35.9 E_r $, $\eta_x=11$,
      $\eta_y=10$ and $\lambda=1000 \ {\rm nm}$.  The kets indicate
      the dominant contribution to the states. Only transversally
      excited bound states couple strongly to the ground trap state as
      displayed by the size of the avoided crossing. The two CIRs
      correspond to COM excitations in either of the transversal
      directions.}
    \label{fig:spectrumplot_ci_sextic_ix6050_iy5000_iz50}
  \end{centering}
\end{figure}

In the following full {\it ab initio} calculations of the spectrum of the general
Hamiltonian \eref{eq:6tic_Hamil_full} are presented to validate the
CRC model. 
The calculations are performed by a full six-dimensional exact-diagonalization
approach which uses a basis of B splines and spherical harmonics \cite{cold:gris09}. 
To incorporate coupling, sextic potentials are used which are an accurate 
representation of a single site of an optical lattice \cite{cold:gris09}. 
The interaction is described by a Born-Oppenheimer potential which can be varied 
to tune the scattering length $a$ to arbitrary values \cite{cold:gris09}.
While in the experiment a quasi 1D trap of
large anisotropy with $\eta_x \approx \eta_y \approx 900$ is used, the
calculations are performed with $\eta_x \approx \eta_y=10$ which is
however already well in the quasi 1D regime \cite{cold:idzi06}. Larger
anisotropies would lead to a prohibitive computational effort.
A fully coupled spectrum for transversal isotropy ($\eta_x=\eta_y$) is
shown in \figref{fig:spectrumplot_ci_sextic_ix5000_iy5000_iz50_zoom2}.
The complex structure of the energy spectrum is highlighting the
importance of the derived selection rules that specify which of the
crossing states couple significantly. The coupling strength is
proportional to the width of the avoided crossings that can be observed
in \figref{fig:spectrumplot_ci_sextic_ix6050_iy5000_iz50} and the lower
graph of \figref{fig:spectrumplot_ci_sextic_ix5000_iy5000_iz50_zoom2}. 
The selection rules are clearly verified: significant avoided crossings appear
only for transversally excited bound states whereas a longitudinal
excitation leads to negligible coupling.

\begin{figure}[ht]
  \begin{centering}
    \includegraphics[width=0.42\textwidth]{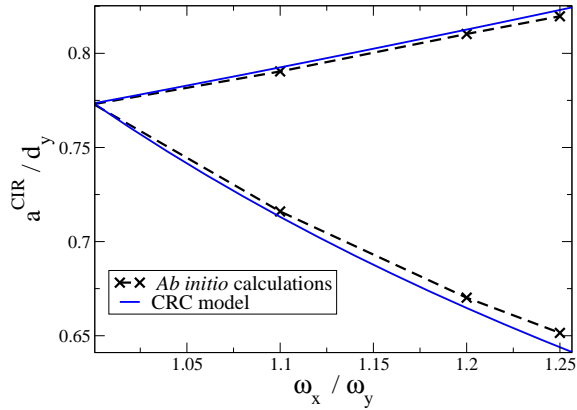}
    \caption{(color online) Positions of inelastic CIRs for different
      values of the transversal anisotropy in quasi 1D. The results of
      the {\it ab initio} calculations are compared to the prediction
      of the CRC model using the same parameters ($^7$Li atoms
      confined in a sextic trapping potential with $V_y = 35.9
      E_r $, $\eta_y=10$ and $\lambda=1000 \ {\rm nm}$) that, however,
      differ from the ones in \figref{fig:cir_splitting_num_formula}.}
    \label{fig:cir_splitting_num_exp}
  \end{centering}
\end{figure}

While in the case of isotropic confinement, $\eta_x = \eta_y$, the transversally excited 
states are degenerate and only a single resonance is visible,
for anisotropic confinement, $\eta_x\neq \eta_y$, this degeneracy is lifted.
In accordance with the CRC model a splitting of the CIR positions appears 
in \figref{fig:spectrumplot_ci_sextic_ix6050_iy5000_iz50}.
In \figref{fig:cir_splitting_num_exp} the CIR positions of the
{\it ab initio} calculations and the CRC model are shown. The stronger
longitudinal confinement results in a systematic shift of the
resonances towards larger values of the scattering length. 
This is mainly caused by the behavior of the
energy of the first trap state. Although this energy could only be
estimated within the CRC model the deviations to the {\it ab initio}
results are less than $1\%$. 
In contrast, the resonance position of the
elastic CIR predicted by the RMH model is independent of the longitudinal confinement.

Next, the CRC model is 
extended to %
2D confinement which is characterized by a single strong transversal
confinement in, e.g., the $x$ direction (i.e.\ $\omega_x \gg \omega_y,\,
\omega_z$).
While \eqref{eq:bound_state_ener} and \eqref{eq:crossing} stay valid,
the contributions of the weakly confined directions to the matrix element in
\eqref{eq:matrix_ele} can be neglected.
Hence, the coupling matrix element between an excited bound state and the
lowest trap state becomes
\begin{align}
  \label{eq:final_select_2D}
   W_\mathbf{n} \approx \delta_{n_y,0} \delta_{n_z,0} 
  \braket{\phi_{n_x} \psi^{(b)} | W_x | \psi_1 \phi_{0} }.
\end{align}
Only a transversally excited bound state with $n_x=2,4,\dots$;
$n_y=n_z=0$ can lead to coupling. Again, only the first excitation
$\mathbf{n} = (2,0,0)$ is dominant,
i.e. only a single resonance appears.
In the experiment \cite{cold:hall10b} a single resonance is observed for 2D
confinement indicated by maximal particle loss at $\frac{a}{d_y}=0.593$
\cite{cold:Haller_priv}.  For the experimental trap parameters the CRC model
predicts the CIR at $\frac{a}{d_y}=0.595$, again, in perfect agreement. This
highlights the fundamental difference to the RMH model which predicts a
resonance for \emph{negative} values of the scattering length in 2D
confinement.
 
As in the 1D case the selection rules of the CRC model are confirmed by full
{\it ab initio} calculations. For $\eta_x = 10$ and $\eta_y=1$ 
both the {\it ab initio} calculations and the CRC model predict the CIR position at
$\frac{a}{d_y}=0.64$ showing again that the resonance position depends also 
on the strength of the weak confinement.


In conclusion, a model describing inelastic CIRs is presented which explains
qualitatively and quantitatively the surprising experimental results of a
splitting of the resonance under transversal anisotropic 1D confinement and a
resonance at positive scattering length in 2D confinement in the loss
experiment \cite{cold:hall10b}. These resonances are caused by the coupling to
bound states with COM excitation in the tightly confined direction. At the
inelastic CIR molecules are formed causing enhanced losses. The previously
known CIRs are elastic and are therefore hardly visible in a loss
experiment. However, they can be observed by measuring the effective
interaction strength directly \cite{cold:froh11}.
The model has important consequences for the experimental realization of a
Tonks-Girardeau gas. As the inelastic-resonance position depends sensitively
on the energy of the resonant excited bound state, it is possible to prepare a
system on the broad elastic CIR where $g_{\rm 1D}\to\infty$, i.e. the
Tonks-Girardeau limit is reached, while being off resonant to the excited
bound state and thus avoiding atom losses \cite{cold:hall09}. In 2D the
elastic and inelastic CIRs are well separated which explains the different
resonance positions observed in \cite{cold:hall10b} and \cite{cold:froh11}.
Since the underlying mechanism of COM-REL coupling leading to an inelastic
CIRs is of very general nature, we believe that they can be observable in
various physical systems. %
\acknowledgments{We are grateful to the \emph{Telekom Stiftung} and
  \emph{Fonds der Chemischen Industrie} for financial support and to
  E. Haller, S. Jochim and G. Shlyapnikov for helpful discussions.}


\begin{thebibliography}{22}%
\makeatletter
\providecommand \@ifxundefined [1]{%
 \@ifx{#1\undefined}
}%
\providecommand \@ifnum [1]{%
 \ifnum #1\expandafter \@firstoftwo
 \else \expandafter \@secondoftwo
 \fi
}%
\providecommand \@ifx [1]{%
 \ifx #1\expandafter \@firstoftwo
 \else \expandafter \@secondoftwo
 \fi
}%
\providecommand \natexlab [1]{#1}%
\providecommand \enquote  [1]{``#1''}%
\providecommand \bibnamefont  [1]{#1}%
\providecommand \bibfnamefont [1]{#1}%
\providecommand \citenamefont [1]{#1}%
\providecommand \href@noop [0]{\@secondoftwo}%
\providecommand \href [0]{\begingroup \@sanitize@url \@href}%
\providecommand \@href[1]{\@@startlink{#1}\@@href}%
\providecommand \@@href[1]{\endgroup#1\@@endlink}%
\providecommand \@sanitize@url [0]{\catcode `\\12\catcode `\$12\catcode
  `\&12\catcode `\#12\catcode `\^12\catcode `\_12\catcode `\%12\relax}%
\providecommand \@@startlink[1]{}%
\providecommand \@@endlink[0]{}%
\providecommand \url  [0]{\begingroup\@sanitize@url \@url }%
\providecommand \@url [1]{\endgroup\@href {#1}{\urlprefix }}%
\providecommand \urlprefix  [0]{URL }%
\providecommand \Eprint [0]{\href }%
\providecommand \doibase [0]{http://dx.doi.org/}%
\providecommand \selectlanguage [0]{\@gobble}%
\providecommand \bibinfo  [0]{\@secondoftwo}%
\providecommand \bibfield  [0]{\@secondoftwo}%
\providecommand \translation [1]{[#1]}%
\providecommand \BibitemOpen [0]{}%
\providecommand \bibitemStop [0]{}%
\providecommand \bibitemNoStop [0]{.\EOS\space}%
\providecommand \EOS [0]{\spacefactor3000\relax}%
\providecommand \BibitemShut  [1]{\csname bibitem#1\endcsname}%
\let\auto@bib@innerbib\@empty
\bibitem [{\citenamefont {Nayak}\ \emph {et~al.}(2008)\citenamefont {Nayak},
  \citenamefont {Simon}, \citenamefont {Stern}, \citenamefont {Freedman},\ and\
  \citenamefont {Das~Sarma}}]{cold:naya08}%
  \BibitemOpen
  \bibfield  {author} {\bibinfo {author} {\bibfnamefont {C.}~\bibnamefont
  {Nayak}}, \bibinfo {author} {\bibfnamefont {S.~H.}\ \bibnamefont {Simon}},
  \bibinfo {author} {\bibfnamefont {A.}~\bibnamefont {Stern}}, \bibinfo
  {author} {\bibfnamefont {M.}~\bibnamefont {Freedman}}, \ and\ \bibinfo
  {author} {\bibfnamefont {S.}~\bibnamefont {Das~Sarma}},\ }\href@noop {}
  {\bibfield  {journal} {\bibinfo  {journal} {Rev.\,Mod.\,Phys.}\ }\textbf
  {\bibinfo {volume} {80}},\ \bibinfo {pages} {1083} (\bibinfo {year}
  {2008})}\BibitemShut {NoStop}%
\bibitem [{\citenamefont {{Haller}}\ \emph {et~al.}(2009)\citenamefont
  {{Haller}}, \citenamefont {{Gustavsson}}, \citenamefont {{Mark}},
  \citenamefont {{Danzl}}, \citenamefont {{Hart}}, \citenamefont {{Pupillo}},\
  and\ \citenamefont {{N{\"a}gerl}}}]{cold:hall09}%
  \BibitemOpen
  \bibfield  {author} {\bibinfo {author} {\bibfnamefont {E.}~\bibnamefont
  {{Haller}}}, \bibinfo {author} {\bibfnamefont {M.}~\bibnamefont
  {{Gustavsson}}}, \bibinfo {author} {\bibfnamefont {M.~J.}\ \bibnamefont
  {{Mark}}}, \bibinfo {author} {\bibfnamefont {J.~G.}\ \bibnamefont {{Danzl}}},
  \bibinfo {author} {\bibfnamefont {R.}~\bibnamefont {{Hart}}}, \bibinfo
  {author} {\bibfnamefont {G.}~\bibnamefont {{Pupillo}}}, \ and\ \bibinfo
  {author} {\bibfnamefont {H.}~\bibnamefont {{N{\"a}gerl}}},\ }\href@noop {}
  {\bibfield  {journal} {\bibinfo  {journal} {Science}\ }\textbf {\bibinfo
  {volume} {325}},\ \bibinfo {pages} {1224} (\bibinfo {year}
  {2009})}\BibitemShut {NoStop}%
\bibitem [{\citenamefont {Paredes}\ \emph {et~al.}(2004)\citenamefont
  {Paredes}, \citenamefont {Widera}, \citenamefont {Murg}, \citenamefont
  {Mandel}, \citenamefont {F{\"o}lling}, \citenamefont {Cirac}, \citenamefont
  {Shlyapnikov}, \citenamefont {H{\"a}nsch},\ and\ \citenamefont
  {Bloch}}]{cold:pare04}%
  \BibitemOpen
  \bibfield  {author} {\bibinfo {author} {\bibfnamefont {B.}~\bibnamefont
  {Paredes}}, \bibinfo {author} {\bibfnamefont {A.}~\bibnamefont {Widera}},
  \bibinfo {author} {\bibfnamefont {V.}~\bibnamefont {Murg}}, \bibinfo {author}
  {\bibfnamefont {O.}~\bibnamefont {Mandel}}, \bibinfo {author} {\bibfnamefont
  {S.}~\bibnamefont {F{\"o}lling}}, \bibinfo {author} {\bibfnamefont
  {I.}~\bibnamefont {Cirac}}, \bibinfo {author} {\bibfnamefont {G.~V.}\
  \bibnamefont {Shlyapnikov}}, \bibinfo {author} {\bibfnamefont {T.~W.}\
  \bibnamefont {H{\"a}nsch}}, \ and\ \bibinfo {author} {\bibfnamefont
  {I.}~\bibnamefont {Bloch}},\ }\href@noop {} {\bibfield  {journal} {\bibinfo
  {journal} {Nature}\ }\textbf {\bibinfo {volume} {429}},\ \bibinfo {pages}
  {277} (\bibinfo {year} {2004})}\BibitemShut {NoStop}%
\bibitem [{\citenamefont {Kinoshita}\ \emph {et~al.}(2004)\citenamefont
  {Kinoshita}, \citenamefont {Wenger},\ and\ \citenamefont
  {Weiss}}]{cold:kino04}%
  \BibitemOpen
  \bibfield  {author} {\bibinfo {author} {\bibfnamefont {T.}~\bibnamefont
  {Kinoshita}}, \bibinfo {author} {\bibfnamefont {T.}~\bibnamefont {Wenger}}, \
  and\ \bibinfo {author} {\bibfnamefont {D.~S.}\ \bibnamefont {Weiss}},\
  }\href@noop {} {\bibfield  {journal} {\bibinfo  {journal} {Science}\ }\textbf
  {\bibinfo {volume} {305}},\ \bibinfo {pages} {1125} (\bibinfo {year}
  {2004})}\BibitemShut {NoStop}%
\bibitem [{\citenamefont {G\"unter}\ \emph {et~al.}(2005)\citenamefont
  {G\"unter}, \citenamefont {St\"oferle}, \citenamefont {Moritz}, \citenamefont
  {K\"ohl},\ and\ \citenamefont {Esslinger}}]{cold:guen05}%
  \BibitemOpen
  \bibfield  {author} {\bibinfo {author} {\bibfnamefont {K.}~\bibnamefont
  {G\"unter}}, \bibinfo {author} {\bibfnamefont {T.}~\bibnamefont
  {St\"oferle}}, \bibinfo {author} {\bibfnamefont {H.}~\bibnamefont {Moritz}},
  \bibinfo {author} {\bibfnamefont {M.}~\bibnamefont {K\"ohl}}, \ and\ \bibinfo
  {author} {\bibfnamefont {T.}~\bibnamefont {Esslinger}},\ }\href@noop {}
  {\bibfield  {journal} {\bibinfo  {journal} {Phys.\,Rev.\,Lett.}\ }\textbf
  {\bibinfo {volume} {95}},\ \bibinfo {pages} {230401} (\bibinfo {year}
  {2005})}\BibitemShut {NoStop}%
\bibitem [{\citenamefont {Haller}\ \emph {et~al.}(2010)\citenamefont {Haller},
  \citenamefont {Mark}, \citenamefont {Hart}, \citenamefont {Danzl},
  \citenamefont {Reich{\-}s{\"o}l{\-}lner}, \citenamefont {Melezhik},
  \citenamefont {Schmelcher},\ and\ \citenamefont {N{\"a}gerl}}]{cold:hall10b}%
  \BibitemOpen
  \bibfield  {author} {\bibinfo {author} {\bibfnamefont {E.}~\bibnamefont
  {Haller}}, \bibinfo {author} {\bibfnamefont {M.~J.}\ \bibnamefont {Mark}},
  \bibinfo {author} {\bibfnamefont {R.}~\bibnamefont {Hart}}, \bibinfo {author}
  {\bibfnamefont {J.~G.}\ \bibnamefont {Danzl}}, \bibinfo {author}
  {\bibfnamefont {L.}~\bibnamefont {Reich{\-}s{\"o}l{\-}lner}}, \bibinfo
  {author} {\bibfnamefont {V.}~\bibnamefont {Melezhik}}, \bibinfo {author}
  {\bibfnamefont {P.}~\bibnamefont {Schmelcher}}, \ and\ \bibinfo {author}
  {\bibfnamefont {H.-C.}\ \bibnamefont {N{\"a}gerl}},\ }\href@noop {}
  {\bibfield  {journal} {\bibinfo  {journal} {Phys.\,Rev.\,Lett.}\ }\textbf
  {\bibinfo {volume} {104}},\ \bibinfo {pages} {153203} (\bibinfo {year}
  {2010})}\BibitemShut {NoStop}%
\bibitem [{\citenamefont {Olshanii}(1998)}]{cold:olsh98}%
  \BibitemOpen
  \bibfield  {author} {\bibinfo {author} {\bibfnamefont {M.}~\bibnamefont
  {Olshanii}},\ }\href@noop {} {\bibfield  {journal} {\bibinfo  {journal}
  {Phys.\,Rev.\,Lett.}\ }\textbf {\bibinfo {volume} {81}},\ \bibinfo {pages}
  {938} (\bibinfo {year} {1998})}\BibitemShut {NoStop}%
\bibitem [{\citenamefont {Girardeau}(1960)}]{cold:gira60}%
  \BibitemOpen
  \bibfield  {author} {\bibinfo {author} {\bibfnamefont {M.}~\bibnamefont
  {Girardeau}},\ }\href@noop {} {\bibfield  {journal} {\bibinfo  {journal}
  {J.\,Math.\,Phys.}\ }\textbf {\bibinfo {volume} {1}},\ \bibinfo {pages} {516}
  (\bibinfo {year} {1960})}\BibitemShut {NoStop}%
\bibitem [{\citenamefont {Lieb}\ and\ \citenamefont
  {Liniger}(1963)}]{cold:lieb63}%
  \BibitemOpen
  \bibfield  {author} {\bibinfo {author} {\bibfnamefont {E.~H.}\ \bibnamefont
  {Lieb}}\ and\ \bibinfo {author} {\bibfnamefont {W.}~\bibnamefont {Liniger}},\
  }\href@noop {} {\bibfield  {journal} {\bibinfo  {journal} {Phys.\,Rev.}\
  }\textbf {\bibinfo {volume} {130}},\ \bibinfo {pages} {1605} (\bibinfo {year}
  {1963})}\BibitemShut {NoStop}%
\bibitem [{\citenamefont {Petrov}\ \emph {et~al.}(2000)\citenamefont {Petrov},
  \citenamefont {Holzmann},\ and\ \citenamefont {Shlyapnikov}}]{cold:petr00}%
  \BibitemOpen
  \bibfield  {author} {\bibinfo {author} {\bibfnamefont {D.~S.}\ \bibnamefont
  {Petrov}}, \bibinfo {author} {\bibfnamefont {M.}~\bibnamefont {Holzmann}}, \
  and\ \bibinfo {author} {\bibfnamefont {G.~V.}\ \bibnamefont {Shlyapnikov}},\
  }\href@noop {} {\bibfield  {journal} {\bibinfo  {journal}
  {Phys.\,Rev.\,Lett.}\ }\textbf {\bibinfo {volume} {84}},\ \bibinfo {pages}
  {2551} (\bibinfo {year} {2000})}\BibitemShut {NoStop}%
\bibitem [{\citenamefont {Bergeman}\ \emph {et~al.}(2003)\citenamefont
  {Bergeman}, \citenamefont {Moore},\ and\ \citenamefont
  {Olshanii}}]{cold:berg03}%
  \BibitemOpen
  \bibfield  {author} {\bibinfo {author} {\bibfnamefont {T.}~\bibnamefont
  {Bergeman}}, \bibinfo {author} {\bibfnamefont {M.~G.}\ \bibnamefont {Moore}},
  \ and\ \bibinfo {author} {\bibfnamefont {M.}~\bibnamefont {Olshanii}},\
  }\href@noop {} {\bibfield  {journal} {\bibinfo  {journal}
  {Phys.\,Rev.\,Lett.}\ }\textbf {\bibinfo {volume} {91}},\ \bibinfo {pages}
  {163201} (\bibinfo {year} {2003})}\BibitemShut {NoStop}%
\bibitem [{\citenamefont {Fr\"ohlich}\ \emph {et~al.}(2011)\citenamefont
  {Fr\"ohlich}, \citenamefont {Feld}, \citenamefont {Vogt}, \citenamefont
  {Koschorreck}, \citenamefont {Zwerger},\ and\ \citenamefont
  {K\"ohl}}]{cold:froh11}%
  \BibitemOpen
  \bibfield  {author} {\bibinfo {author} {\bibfnamefont {B.}~\bibnamefont
  {Fr\"ohlich}}, \bibinfo {author} {\bibfnamefont {M.}~\bibnamefont {Feld}},
  \bibinfo {author} {\bibfnamefont {E.}~\bibnamefont {Vogt}}, \bibinfo {author}
  {\bibfnamefont {M.}~\bibnamefont {Koschorreck}}, \bibinfo {author}
  {\bibfnamefont {W.}~\bibnamefont {Zwerger}}, \ and\ \bibinfo {author}
  {\bibfnamefont {M.}~\bibnamefont {K\"ohl}},\ }\href@noop {} {\bibfield
  {journal} {\bibinfo  {journal} {Phys.\,Rev.\,Lett.}\ }\textbf {\bibinfo
  {volume} {106}},\ \bibinfo {pages} {105301} (\bibinfo {year}
  {2011})}\BibitemShut {NoStop}%
\bibitem [{\citenamefont {Peng}\ \emph {et~al.}(2010)\citenamefont {Peng},
  \citenamefont {Bohloul}, \citenamefont {Liu}, \citenamefont {Hu},\ and\
  \citenamefont {Drummond}}]{cold:peng10}%
  \BibitemOpen
  \bibfield  {author} {\bibinfo {author} {\bibfnamefont {S.-G.}\ \bibnamefont
  {Peng}}, \bibinfo {author} {\bibfnamefont {S.~S.}\ \bibnamefont {Bohloul}},
  \bibinfo {author} {\bibfnamefont {X.-J.}\ \bibnamefont {Liu}}, \bibinfo
  {author} {\bibfnamefont {H.}~\bibnamefont {Hu}}, \ and\ \bibinfo {author}
  {\bibfnamefont {P.~D.}\ \bibnamefont {Drummond}},\ }\href@noop {} {\bibfield
  {journal} {\bibinfo  {journal} {Phys.\,Rev.\,A}\ }\textbf {\bibinfo {volume}
  {82}},\ \bibinfo {pages} {063633} (\bibinfo {year} {2010})}\BibitemShut
  {NoStop}%
\bibitem [{\citenamefont {Zhang}\ and\ \citenamefont
  {Zhang}(2011)}]{cold:zhan11}%
  \BibitemOpen
  \bibfield  {author} {\bibinfo {author} {\bibfnamefont {W.}~\bibnamefont
  {Zhang}}\ and\ \bibinfo {author} {\bibfnamefont {P.}~\bibnamefont {Zhang}},\
  }\href@noop {} {\bibfield  {journal} {\bibinfo  {journal} {Phys.\,Rev.\,A}\
  }\textbf {\bibinfo {volume} {83}},\ \bibinfo {pages} {053615} (\bibinfo
  {year} {2011})}\BibitemShut {NoStop}%
\bibitem [{\citenamefont {Busch}\ \emph {et~al.}(1998)\citenamefont {Busch},
  \citenamefont {Englert}, \citenamefont {Rzazewski},\ and\ \citenamefont
  {Wilkens}}]{cold:busc98}%
  \BibitemOpen
  \bibfield  {author} {\bibinfo {author} {\bibfnamefont {T.}~\bibnamefont
  {Busch}}, \bibinfo {author} {\bibfnamefont {B.-G.}\ \bibnamefont {Englert}},
  \bibinfo {author} {\bibfnamefont {K.}~\bibnamefont {Rzazewski}}, \ and\
  \bibinfo {author} {\bibfnamefont {M.}~\bibnamefont {Wilkens}},\ }\href@noop
  {} {\bibfield  {journal} {\bibinfo  {journal} {Found.\,Phys.}\ }\textbf
  {\bibinfo {volume} {28}},\ \bibinfo {pages} {549} (\bibinfo {year}
  {1998})}\BibitemShut {NoStop}%
\bibitem [{\citenamefont {Idziaszek}\ and\ \citenamefont
  {Calarco}(2006)}]{cold:idzi06}%
  \BibitemOpen
  \bibfield  {author} {\bibinfo {author} {\bibfnamefont {Z.}~\bibnamefont
  {Idziaszek}}\ and\ \bibinfo {author} {\bibfnamefont {T.}~\bibnamefont
  {Calarco}},\ }\href@noop {} {\bibfield  {journal} {\bibinfo  {journal}
  {Phys.\,Rev.\,A}\ }\textbf {\bibinfo {volume} {74}},\ \bibinfo {pages}
  {022712} (\bibinfo {year} {2006})}\BibitemShut {NoStop}%
\bibitem [{Note1()}]{Note1}%
  \BibitemOpen
  \bibinfo {note} {The shifted bound state shown in Fig.1 in \cite
  {cold:hall10b} and Fig.2 in \cite {cold:berg03} does, in fact, not exist in
  the energy spectrum of the \protect \emph {full} relative motion Hamiltonian,
  see corresponding figures in \cite {cold:busc98,*cold:idzi06}, but of a
  projected Hamiltonian.}\BibitemShut {Stop}%
\bibitem [{\citenamefont {Liang}\ and\ \citenamefont
  {Zhang}(2008)}]{cold:lian08}%
  \BibitemOpen
  \bibfield  {author} {\bibinfo {author} {\bibfnamefont {J.-J.}\ \bibnamefont
  {Liang}}\ and\ \bibinfo {author} {\bibfnamefont {C.}~\bibnamefont {Zhang}},\
  }\href@noop {} {\bibfield  {journal} {\bibinfo  {journal} {Phys.\,Scr.}\
  }\textbf {\bibinfo {volume} {77}},\ \bibinfo {pages} {025302} (\bibinfo
  {year} {2008})}\BibitemShut {NoStop}%
\bibitem [{Note2()}]{Note2}%
  \BibitemOpen
  \bibinfo {note} {It turns out that this approximation is exact at the
  resonance position in the case of transversal isotropic, harmonic
  confinement.}\BibitemShut {Stop}%
\bibitem [{Note3()}]{Note3}%
  \BibitemOpen
  \bibinfo {note} {The maximal loss positions are determined by shifting the
  known ``edge'' positions by a constant offset $\Delta a = 89\ a_0$ that is
  obtained from the isotropic case.}\BibitemShut {Stop}%
\bibitem [{\citenamefont {Grishkevich}\ and\ \citenamefont
  {Saenz}(2009)}]{cold:gris09}%
  \BibitemOpen
  \bibfield  {author} {\bibinfo {author} {\bibfnamefont {S.}~\bibnamefont
  {Grishkevich}}\ and\ \bibinfo {author} {\bibfnamefont {A.}~\bibnamefont
  {Saenz}},\ }\href@noop {} {\bibfield  {journal} {\bibinfo  {journal}
  {Phys.\,Rev.\,A}\ }\textbf {\bibinfo {volume} {80}},\ \bibinfo {pages}
  {013403} (\bibinfo {year} {2009})}\BibitemShut {NoStop}%
\bibitem [{\citenamefont {{E. Haller}}()}]{cold:Haller_priv}%
  \BibitemOpen
  \bibfield  {author} {\bibinfo {author} {\bibnamefont {{E. Haller}}},\
  }\href@noop {} {}\bibinfo {note} {{private communication}}\BibitemShut
  {NoStop}%
\end{thebibliography}

%

\end{document}